\documentclass[12pt]{article}
\title{Extremal black disks in QCD}
\author{Alexey V. Popov\thanks{email: avp@novgorod.net} \\ \\ 
\small{\emph{Novgorod State University, B. S.-Peterburgskaya Street 41,}}\\
\small{\emph{Novgorod the Great, Russia, 173003}}
} 
\date{}
\begin{document}
\maketitle
\abstract{
We argue that in the high energy QCD a true black disk wave function 
necessarily contains many quarks.
This corresponds to necessity of non-vacuum reggeon loops in formation of a black disk. 
The result comes from decomposition of the black disk S-matrix in characters on group manifold. 
}
\newcommand{\ket}[1]{\ensuremath{|#1\rangle}}
\newcommand{\bra}[1]{\ensuremath{\langle#1|}}
\newcommand{\braket}[2]{\ensuremath{\langle#1|#2\rangle}}
\section*{}
Usually black disk defined as some dense state, which absorbs any projectile. In this paper we consider a reverse case where black disk
is the fast projectile and scatters in
an external color field.  
Any hardon in the QCD can be described by a wave function represented as certain vector in Hilbert space.
Consider a collision of a black disk state $\ket{\Psi}$ with some other state named as target. We take 
small black disk and very large and dense target. 
We suppose that target color field much stronger than projectile field. This does not suppose that target is a black disk too.
Key issue here is that, freely choosing target, we can construct an external color field as we want.
%The target is not necessary black disk but only one requirement that
%the classical fields generated by target must be much stronger then fields generated by black disk. 
In the considered case the projectile elastic S-matrix has form
\begin{equation}
S=\int \bra{\Psi} e^{i\int \alpha_a(x)\rho^a(x) dx} \ket{\Psi} W[\alpha] D\alpha
\end{equation}
where $\alpha_a(x)$ is a classical field generated by some quasiclassical element of target wave function, $\rho^a(x)$ is the operator of color charge 
(generator of gauge transformations) which acts on the projectile state, 
$x$ is a transverse position. Functional $W[\alpha]$ is a weight functional obtained by averaging on target wave function.
Since the target is arbitrary, we shall not perform average on $\alpha$ and consider fixed $\alpha$ configuration.
So the elastic projectile S-matrix becomes to a functional S[$\alpha]$.

In the previous speculations there is one mistake. Though fields $\alpha_a(x)$ can be very large, but due to compactness of a gauge
group scattering amplitude is limited. So in general we have no way to compare fields $\alpha_a(x)$ in the context of scattering.
Nevertheless we define black disk state as state which S-matrix is zero if $\alpha_a(x)\neq 0$ in the area where black disk located 
and is 1 if $\alpha_a(x)=0$ in the same region.  
We shall name this state as extremal (or true) black disk because 
usually in the current literature more weak definitions are used.
We define
\begin{equation} \label{f_2}
S[\alpha]=\bra{\Psi} e^{i\int \alpha_a(x)\rho^a(x) dx} \ket{\Psi}
\end{equation}
where it should be stated that $x$-integral in (\ref{f_2}) is limited by points where 
$\rho^a(x)\ket{\Psi}\neq 0$.
Since a true black disk must be black for any target and any $W[\alpha]$, we must claim
\begin{equation} \label{eq_1}
S[\alpha]=Z\prod_{x,a} \delta(\alpha_a(x))
\end{equation}
in sense of functional generalization of the $\delta$-function. 
The factor $Z$ is the normalization constant which is not important for us.
It was proposed in \cite{Kovner_06} to use variables $\omega^a(x)$ which
arises from Fourier transform of $S[\alpha]$. 
\begin{equation}
S[\omega]=\int S[\alpha] e^{i\int \alpha_a(x) \omega^a(x) dx} D\alpha
\end{equation}
Then for infinite black disk we obtain constant functional. Variables $\omega^a(x)$ usually refer to as a classical color charge.
But this method applicable only for small $\alpha_a$. For arbitrary $\alpha_a$ we should not forget about $SU(N_c)$ compactness. 
Via exponential map $e^{i\alpha_a \rho^a}$ (a map from a lie algebra to some representation) variables $\alpha_a(x)$ can be viewed as canonical 
coordinates on group manifold.  Integration on $\alpha_a(x)$ must be considered as integration on group manifold with Haar
measure\footnote{For diagonal Killing form $Sp(T^aT^b)\sim\delta^{ab}$ and small $\alpha_a$ Haar measure degenerate to usual $\prod_a d\alpha_a$.}. 
Instead of $\alpha_a(x)$ we introduce variables $g(x)$ which is map from the transverse plane to the group manifold $g(x): R^2\rightarrow G$.
Claim (\ref{eq_1}) can be rewritten as
\begin{equation}
S[g]=\prod_x \delta(g(x)-e)
\end{equation}
where $e$ is the identity element of group $G$. Function $\delta(g-e)$ can be viewed as function on group manifold with localized supremum.
Expression $g-e$ is rather formal and has sense only in coordinates. 
In general we define $\delta$-function as the linear functional over functions on group manifold: 
$\int \delta(g-e) f(g) dg = f(e)$

On a compact group instead of using Fourier transform we must use matrix elements of irreducible representations as complete basic for
space of functions. Well known Peter-Weyl theorem says that functions
\begin{equation}
b^R_{ij}(g)=\sqrt{d_R} \pi^R_{ij}(g)
\end{equation}
forms a orthonormal and complete basic in a Hilbert space $L^2(G,dg)$.
\begin{equation}
\int \overline{b^Q_{kl}(g)} b^R_{ij}(g) dg=\delta_{RQ} \delta_{ki} \delta_{lj}
\end{equation}
Letters $R$,$Q$ denotes irreducible representations, $d_R$ is a dimension of corresponding representation, $\pi^R_{ij}(g)$ is a matrix 
element of a representation $R$ of a group element $g$.

Function $f(g)$ is called central function (or class function) if $f(hgh^{-1})=f(g)$ for any $h\in G$. It is easy to show that the 
function $\delta(g-e)$ is central due to invariance of Haar measure
\begin{equation}
\int \delta(hgh^{-1}-e) f(g) dg=\int \delta(g-e) f(h^{-1}gh) dg=f(e)
\end{equation}
Direct consequence of the Peter-Weyl theorem is that characters of irreducible representations forms a orthonormal and complete basic 
in a Hilbert space of central functions. Characters is given by
\begin{equation} \label{eq_4}
\chi_R(g)=\sum_i \pi^R_{ii}(g)
\end{equation}
\begin{equation}
\int \overline{\chi_R(g)} \chi_Q(g) dg=\delta_{RQ}
\end{equation}
Any central function $f(g)$ can be decomposed in a series on characters 
\begin{equation}
f(g)=\sum_R f_R\chi_R(g)
\end{equation}
\begin{equation}
f_R=\int \overline{\chi_R(g)}f(g) dg
\end{equation}
Here we are interested decomposition of the $\delta$-function.
\begin{equation} \label{eq_2}
\delta(g-e)=\sum_R d_R \chi_R(g)
\end{equation}
where we used the obvious fact that $\chi_R(e)=d_R$. It should be emphasized that in case of gauge group $SU(N_c)$ and $N_c>2$ there are
representations which is not equivalent to its complex conjugate. For example quark is not equivalent representation for antiquark.  
Such representations have complex characters, but it is not pure imaginary near $e$. The formula (\ref{eq_2}) is rather 
mathematical idealization. A infinite series is irrelevant in physical sense. In the real world the $\delta$-function must be smeared
near $e$ and so the black disk S-matrix is not zero for very small fields $\alpha_a$. This equivalent to some truncation of series (\ref{eq_2})
which, at first sight, is  hardly divergent due to $d_R$ growing. The convergence of (\ref{eq_2}) is achieved by alternating characters
values at non singular points.
High representations in (\ref{eq_2}) corresponds to high frequencies in usual Fourier analysis. This happens due to growing of
the second Casimir operator $D_2$ with dimension of representation \cite{Jeon_04}. Second Casimir operator on functions is 
Laplace operator. Eigenvalues of Laplace operator correspond to square of frequencies vector in a flat case. 
It well known from wave packets calculations that high frequencies suppression 
equivalents to spreading of a wave packet and in order to obtain  more narrow packed we must include higher harmonics.
For example, characters
for group $SU(2)$ can be calculated explicitly
\begin{equation}
\chi_l(t)=\sum_{m=-l}^{l} e^{imt}=\frac{\sin(l+\frac{1}{2})t}{\sin{\frac{t}{2}}}
\end{equation}
where $t$ is a coordinate on a diagonal matrixes. 

Characters has useful properties. 
Cyclicity of the trace in (\ref{eq_4}) gives $\chi(hg)=\chi(gh)$.
Group multiplication induces two kinds of global vector field, so called
left and right invariant vector fields $J_+^a$ and $J_-^a$. Namely, there is a linear map from lie algebra to vector fields on group manifold.
Acting on characters, invariant vector fields gives equal results.
\begin{equation} \label{eq_3}
J_+^a\chi(g)= \left. \frac{\partial}{\partial h^a}\chi(hg)\right|_{h=0}=
\left. \frac{\partial}{\partial h^a}\chi(gh)\right|_{h=0}=J_-^a\chi(g)
\end{equation}
From unitarity of representations we have
\begin{equation}
\left.J^a_\pm \chi(g) \right|_{g=e}=0
\end{equation} 
The functional (\ref{eq_1}) obeys JIMWLK equation \cite{Kovner_06}.
The JIMWLK evolution equation can be expressed in terms of invariant vector field \cite{Weigert_05}. 
\begin{equation}
\frac{dS[\alpha]}{dY}=\int\limits_{zxy}^{\phantom{z}} K_{xyz}
\left( -J_+^a(x)J_+^a(y) -J_-^a(x)J_-^a(y)+2V_{ba}(z)J_+^b(x)J_-^a(y)  \right) S[\alpha]
\end{equation}
\begin{equation}
K_{xyz}=\frac{g^2}{(2\pi)^3}\frac{(\vec z-\vec y)(\vec z-\vec x)}{(\vec z-\vec y)^2(\vec z-\vec x)^2}
\end{equation}
where $V_{ba}(z)=\exp (i\alpha_c(z)T^c)_{ba}$ is a gluon scattering amplitude in an external field $\alpha_a$. If we take black disk functional (\ref{eq_1}) then 
using property (\ref{eq_3}) 
we can replace $J_\pm^a(x)\rightarrow J^a(x)$. Due to smoothness of $V_{ba}(z)$ as functional of $\alpha$ and
localization of the $\delta$-function near $e$, we have $V_{ba}(z)S[\alpha]\rightarrow V_{ba}|_{\alpha=0}S[\alpha] \rightarrow  \delta_{ba}S[\alpha]$.
Finally, it is straightforward to see that $dS[\alpha]/dY=0$. Strictly speaking the JIMWLK equation is only valid in dilute regime, so
our calculation have only demonstrative purposes. 

The JIMWLK Hamiltonian is a linear operator. If we have its eigenvalue $-\lambda$  then at high $Y$ corresponded 
eigenfunctional disappears as $\exp(-\lambda Y)$. A true black disk has zero eigenvalue. So it is a stable state of the evolution.
This is not true at the black disk border where there is a transition region (crossover) to the white state. Obviously that
black disk must grow when we go to a high $Y$. Hence at the black disk border an interesting process will occur. Emitted quarks and gluons 
will saturate the border by higher characters thus it will become black.

Next questions is about possible partonic interpretation of characters $\chi_R$. Most obvious method is using density matrix. Consider
one parton with color charge $R$. Lets the parton color space density matrix be proportional to 1 (maximum entropy state).
Normalization factor is simple $d_R^{-1}$. Then the elastic scattering amplitude in a external field $\alpha_a$ is 
\begin{equation}
\langle e^{i\alpha_a T^a_R}\rangle=\frac{1}{d_R}Sp\left(e^{i\alpha_a T^a_R}\right)=\frac{1}{d_R}\chi_R(g)
\end{equation}
where averaging is over density matrix.
The average color charge within black disk is zero.
\begin{equation}
\langle\rho^a\rangle=\left.J^a \delta(g-e) \right|_{g=e} =0
\end{equation}
This behavior corresponds to McLerran-Venugopalan(MV) states where in addition required vanishing of $\langle\rho^a(x)\rho^b(y)\rangle$ for $x\neq y$.
In difference from MV model in our black disk state the averaged charge square  $\rho^a(x) \rho^a(x)$ is divergent.

It follows from (\ref{eq_2}) that it is absolutely necessary to have many quarks and antiquarks in a black disk wave function. Many representations can not be
obtained via reducing of tensor products of gluon representation. For example in the $SU(3)$ case such representations are 
$\textbf 3,\bar\textbf 3,\textbf 6 \ldots$ At this time we do not able to predict that such scenario actually can be realized in nature
but necessity of existence of partons in the fundamental representation in a \textbf{true} black disk state was proven. 
So we can suggest the following hypothetical scenario. Starting from dilute projectile such as dipole, when boosting it to high $Y$ we predominantly 
generate many new gluons in a wave function. When the gluons density to become high then we have large 
probability to emit soft quarks into new opened phase space. This corresponds to mechanism of non-vacuum reggeon in QCD.
See \cite{Kovchegov_03} for modern example of a quark propagation in rapidity.
Usually such processes are kinematically suppressed in comparison with a gluon emission but we can expect that in case of high number
of sources it can be very probable. Fast gluon converts into fast quark and slow antiquark and vice versa. 
That fact that in order to obtain appropriate characters we took maximum entropy density matrix naturally correspond to 
expected chaotic behavior of a wave function boosted to high energy. Another argument comes from \cite{Kovner_06} 
where it has been specified that there are only two vacuum states exists in the high energy QCD evolution. 
These states are white and black disks. So investigated extremal black disk state can pretend to be a black vacuum of full QCD evolution.

\bigskip \noindent \textbf{Acknowledgment:} We thank N.V. Prikhod'ko for feedback and useful remarks.

\end{document}